\documentclass[conference]{IEEEtran}
\IEEEoverridecommandlockouts

\usepackage[hmargin=.75in,vmargin=1in]{geometry}
\usepackage[american]{babel}
\usepackage[T1]{fontenc}
\usepackage{times}
\usepackage{caption}
\usepackage[usenames, dvipsnames]{color}
\usepackage[utf8]{inputenc}
\usepackage{lmodern}
\usepackage{caption}
\usepackage{tabularx}
\usepackage{subcaption}
\usepackage{relsize}
\usepackage{float}
\usepackage{cite}
\usepackage{multirow}
\usepackage{multicol}
\usepackage{makecell}
\usepackage{amsmath,amssymb,amsfonts}
\usepackage{algorithmic}
\usepackage{graphicx}
\usepackage{textcomp}
\usepackage{arydshln}
\usepackage{tikz}
\usetikzlibrary{positioning,fit,calc}
\usepackage{amsmath}

\usepackage{textcomp}
\usepackage{epsfig,graphicx}
\usepackage{xcolor}
\usepackage{amsfonts,amsmath,amssymb}
\usepackage{fixltx2e} % Fixing numbering problem when using figure/table* 
\usepackage{booktabs}

\def\checkmark{\tikz\fill[scale=0.4](0,.35) -- (.25,0) -- (1,.7) -- (.25,.15) -- cycle;} 
\usetikzlibrary{positioning}

\restylefloat{table}
\newcolumntype{C}{>{\small\centering\arraybackslash}X}

\def\BibTeX{{\rm B\kern-.05em{\sc i\kern-.025em b}\kern-.08em
    T\kern-.1667em\lower.7ex\hbox{E}\kern-.125emX}}
    
\begin{document}

\title{What we learn from learning - Understanding capabilities and limitations of machine learning in botnet attacks \\
\thanks{David Santana is with the University of North Carolina at Greensboro, Greensboro, USA (email: d\_santan@uncg.edu).}
\thanks{
Shan Suthaharan is with the University of North Carolina at Greensboro, Greensboro, USA (email: s\_suthah@uncg.edu )}
\thanks{
Somya D. Mohanty is with the University of North Carolina at Greensboro, Greensboro, USA (email: sdmohant@uncg.edu)} 
}

% \\

% \author{\IEEEauthorblockN{1\textsuperscript{st} David Santana, Cory Sabol}
% \IEEEauthorblockA{\textit{Computer Science. University of North Carolina at Greensboro} \\
% Greensboro, US \\
% d_santana@uncg.edu, cssabol@uncg.edu}
% \and
% \IEEEauthorblockN{2\textsuperscript{nd} Shan Suthaharan}
% \and
% \IEEEauthorblockN{3\textsuperscript{rd} Somya Mohanty}

\author{
{\bfseries David Santana, Shan Suthaharan, and Somya Mohanty}
% \\ $^2$Department Name, Company Name / Institution Name, City, State, Country\\
}

\maketitle

\begin{abstract}
With a growing increase in botnet attacks, computer networks are constantly under threat from attacks that cripple cyber-infrastructure. Detecting these attacks in real-time proves to be a difficult and resource intensive task. One of the pertinent methods to detect such attacks is signature based detection using machine learning models. This paper explores the efficacy of these models at detecting botnet attacks, using data captured from large-scale network attacks. Our study provides a comprehensive overview of performance characteristics two machine learning models --- Random Forest and Multi-Layer Perceptron (Deep Learning) in such attack scenarios. Using Big Data analytics, the study explores the advantages, limitations, model/feature parameters, and overall performance of using machine learning in botnet attacks / communication. With insights gained from the analysis, this work recommends algorithms/models for specific attacks of botnets instead of a generalized model.  
\end{abstract}

\noindent\textbf{Keywords:}
 {\small  
Botnet detection, Network Security, Machine Learning, Ensemble Methods, Deep-Learning
 } %%%% Replace with your keywords

\section{Introduction}
In an increasingly connected world of machines where usability takes presence over security, computers/machines have become more vulnerable to exploits by malicious entities. A large portion of such machines are utilized for their computational and connectivity resources to conduct synchronized attacks across networks. Botnets are a  group of exploited machines (with malicious software) that are leveraged for large scale attacks across different cyber-infrastructure components \cite{alexander_k}. 

Attacks from botnets include  email/message based spam, deploying malware, executing distributed denial of service (DDoS) attacks, and even spreading the botnet software to more computers \cite{wang2010peer}. In Q1 2017 \cite{alexander_k}, cyber-infrastructure resources in almost 72 countries were targeted using botnets. The Federal Bureau of Investigation \cite{botnetfbi} reports that 18 computers per second are participating in botnets, amounting to 500 million computers worldwide. With growing participation (maliciously exploited) of connected devices in the form of Internet-Of-Things (IoT) in botnets \cite{kolias2017ddos}, detection of such attacks have been key towards securing infrastructure.  

While machine learning \cite{jyothsna2011review} and network modelling techniques \cite{garcia2009anomaly} have been widely used towards detection of botnet attacks, most of the approaches are tailored towards detection of specific types of such attacks. The paper explores the advantages and disadvantages of such models in a real-world scenario. More specifically, utilizing real network data captured across a computer network \cite{Garc_a_2014}, for seven different types of botnet attacks, the paper examines the utility of different machine learning models in detecting such attacks. 

Using information collected from Netflow records \cite{netflow}, the approach examines the models across standard performance/accuracy measures of - Precision, Recall, and F1-Measure for \textit{attack} versus \textit{non-attack} records using supervised models. The analysis removes any device/user identifiable information from the records by using aggregate statistics and does not perform any deep-packet payload inspection. The limitations were purposely put in place for developing efficient models capable of operating in real-time scenario with limited resources (such as routers).

This study also aims to identify the ideal parameters of detection of specific attacks by conducting exhaustive search of parameters and features for the machine learning models. Big Data techniques were utilized to compute the performance characteristics of the models using parallelization techniques. The goal is to improve detection of botnet attacks in real-world real-time systems, where intrusion detection systems (IDS) can be pro-active utilizing the appropriate models for detection of attacks and eventual mitigation.

\section{Background}
\label{background}

The most common model for operating botnets is the command and control (C\&C) architecture \cite{zeidanloo2009botnet}. The controller is a master who issues commands for the distributed architecture of bots through legitimate Internet Relay Chat (IRC) channels. The predominant detection methods for botnets are based on analyzing communication/traffic streams. They include statistical approaches, where anomaly detection \cite{ gu2008botsniffer} is utilized for identifying abnormal activity related to botnets. More recently, machine learning methods \cite{stevanovic2014efficient} have been used to classify network traffic (normal/attack). Livadas \textit{et al.} \cite{livadas2006usilng} utilized multi-stage Bayesian network classifiers to detect botnets, where the first stage is utilized to characterize IRC traffic, and the second for botnet communication. Saad \textit{et. al.} \cite{5971980} compared the performance of five different supervised machine learning algorithms to detect P2P botnet attacks. The BClus methods proposed by Garcia \textit{et. al.} \cite{garcia2014empirical}, are a unique approach towards utilizing unsupervised clustering algorithms to annotate botnet attacks. Supervised machine-learning approaches such Random Forest \cite{zhang2008random} and Deep-Learning \cite{javaid2016deep} models have been explored by various studies establishing their effectiveness in analyzing network based communication/traffic. However, a comparative analysis on the advantages and disadvantages of each of the approaches has been lacking.

% Apart from the centralized C\&C architecture, recent models \cite{zhu2008botnet} use peer-to-peer (P2P) communication protocols where bots are capable of talking to each other and conducting distributed attacks. While the P2P model mitigates the single point of failure in the C\&C architecture, it does have latency issues which arise form distributed communication. Many models circumvent detection, by using HTTP based communication \cite{li2009botnet} instead of the standard IRC channel. 

\begin{table}[ht!]
    \centering
    \scalebox{0.5}{
    \begin{tabular}{|c|c|c|c|c|c|c|c|c|c|c|c|} 
    \hline
         Id & \makecell{Time \\ (hrs)} & \#Packets & \#Netflows & Size & Bot & \#Bots & IRC & SPAM & DDOS \\ \hline
         1 & 6.15   & 71.9 M    & 2 M    & 52GB       & Neris     & 1    & \checkmark & \checkmark &  \\ \hline
         2 & 4.21   & 71.8 M    & 1.8 M     & 60GB       & Neris     & 1    & \checkmark & \checkmark & \\ \hline
         3 & 66.85  & 167.7 M   & 4.7 M    & 121GB      & Rbot      & 1    & \checkmark & & \\ \hline
         4 & 4.21   & 62 M    & 1.1 M    & 53GB       & Rbot      & 1    & \checkmark & & \checkmark \\ \hline
         5 & 11.63  & 4.4 M     & 1.2 M      & 37.6GB     & Viruit    & 1    & & \checkmark &\\ \hline
         6 & 5.18   & 115.4 M   & 2.7 M    & 94GB       & Neris     & 10   & \checkmark & \checkmark & \\ \hline
         7 & 4.75  & 90.3 M    & 1.3 M    & 73GB       & Rbot      & 10   & \checkmark & & \checkmark\\ \hline
         8 & 0.26  & 6.3 M     & 107 K      & 5.2GB      & Rbot      & 3    & \checkmark & & \checkmark\\ \hline
         9 & 16.36 & 50.8 M    & 1.9 M    & 34GB       & Virut     & 1    & & \checkmark & \\ \hline
         
    \end{tabular}
    }
    \caption{Botnet Data - Netflow records of communication}
    \label{tab:binetinfo}
\end{table}

% \begin{figure*}[ht!]
%     \centering
%     \includegraphics[height=2in]{./images/botnet_scenarios.jpg}
%     \caption{Data in each botnet scenario.}
%     \label{fig:ctu-13types}
% \end{figure*}
% 
% \begin{figure*}[ht!]
%     \centering
%     \includegraphics[height=1.6in]{./images/botnet_char.png}
%     \caption{Characteristics of Botnet Attacks}
%     \label{fig:botnet_char}
% \end{figure*}

A key hurdle in evaluating the capabilities of detection models is the availability of real-world datasets. While the NSL-KDD (KDDCUP-99) \cite{tavallaee2009detailed} has provided researchers with labeled data, including Denial Of Service (DOS), unauthorized access, remote access, and probing attacks, it does not represent the different varieties of attacks current botnets are capable of. 

The CTU-13 dataset \cite{garcia2014empirical} is a comprehensive dataset representing different scenarios of botnet attacks. The CTU-13 dataset, was developed for a comparative measure of detection performance among various algorithms. The dataset contains real-world network capture of thirteen different types of attacks conducted using different botnets (Neris, Rbot, Virut, etc.). The dataset is in the form of bidirectional netflow records capturing ---  Start Time, End Time, Duration, Source IP address, Source Port, Direction, Destination IP address, Destination Port, State, SToS, Total Packets and Total Bytes of each connection in the network. 

For comparative measure of detection performance in different scenarios of botnet attacks, the records have been labeled with --- Normal, Background, C\&C, and Botnet flows. The dataset represents a multitude of characteristics of botnet attacks and their communications. Table \ref{tab:binetinfo}, denotes the different types of communication and attack characteristics which covers SPAM and DDoS attacks, and IRC communication. As a result the dataset provides for an excellent opportunity to study and compare the abilities of different machine learning models in detecting the different characteristics of the attacks.

\section{Approach}
\label{approach}
Utilization of machine learning models in detecting botnet operation in a network can be categorized into two generalized approaches --- 1) Payload, and 2) Traffic based. As the name suggests, the payload based approach trains models based on features extracted from the payload / data component of the packets transmitted across the network. The drawbacks of such a models are the resource intensive nature (where every packet has to be analyzed for features), issues with privacy, and encrypted data (features cannot be extracted). The traffic based approach aims at alleviating some of the drawbacks of the model by analyzing the packet headers or Netflow records of communication. While privacy is still a concern with such an approach (such as individual IP addresses in features), this can be mitigated by using aggregation of records for time windows. 

% \begin{figure}
%   \centering
%   \includegraphics[height=1.4in]{./images/botnet_labels.jpg}
%   \caption{Botnet labels on each connection}
%   \label{dataset_info}
% \end{figure}

Our approach utilizes the temporal domain of the Netflow records by developing features for time-windows in a network. We develop aggregated features for time durations/windows which are then utilized to train machine-learning models. For each dataset, the features of the aggregated Netflows are calculated by taking into account all Netflows that were initiated or were continuing (initiated in a earlier window but currently active) in a particular window. The values of each Netflow record which falls into either of the two categories were taken into account for developing features of the window. 

The 13 datasets were merged into 4 different categories based on the category of attack/communication --- 1) DDoS (4, 7, 8), 2) Spam (1,2,5,6,9), and 3) IRC (1,2,3,4,6,7,8). The target label for each of the records of the dataset were developed in the same way as features for the time-window, considering Botnet and C\&C records as \textit{'attack'} labels and only the normal records as \textit{non-attack} label. In other words, if a single Netflow record in the time-window has Botnet and/or C\&C tags, the entire window is considered as an attack window. If no records had those tags in the data, the window is considered non-attack. Further more, multiple datasets were created (from the merged) data using different window sizes ($0.01$, $1.0$, $10$, $30$, $60$ seconds) for aggregation and evaluation of the models. 

This dataset did not use any records in the original dataset tagged as 'background'. The goal here was to evaluate the efficiency of the developed models on verified records which have been identified by subject matter expert (SME). With the best developed models (using feature and parameter tuning) in such an scenario, the effectiveness of the models can then be tested in an environment where a large portion of the data is unverified. For this propose, a secondary dataset was developed which considered background tag into developing the time-window based features and labeling. In this case, the background tag was assumed to be 'non-attack' traffic considering all of the attack traffic was labeled by the SME in the environment.

% Labeling of the windows is done by considering Botnet and C\&C labels in the records to be attacks, and only Normal to be non-attack traffic. In other words, if a single Netflow record in the time-window has Botnet and/or C\&C labels, the entire window is considered as an attack window. Multiple duration time-slices were tested for their accuracy and performance (described in later sections).

\subsection{Feature Extraction}

The identified features of the study can broadly be classified into two categories --- 1) Extracted , and 2) Analyzed features. The extracted features record simplistic observations of connections within a time-window. Most of the features in this category are general counts ($\mathcal{N}$) or mean ($\mathcal{\bar{W}}$) values are aggregated over connections within a time-window. Table \ref{tab:features}, describes the various categories of such features, where observations of 1) Connection Information, 2) IP Address, 3) Port, and 4) Protocols used by the Netflow records are used. Extracted features in the connection category record counts of different types of connection and mean duration (\textit{$\mathcal{\bar{W}}$\_duration}) for each window. In the IP address category, aggregated counts of different classes of source and destination IP address classes are kept. Similarly, port variables record count of utilization of different source and destination ports above and below 1024 (below 1024 are well-known ports). Protocols used by the connects are also kept track of in the variables representing counts of TCP (\textit{$\mathcal{N}$\_TCP}), UDP (\textit{$\mathcal{N}$\_UDP}), and ICMP (\textit{$\mathcal{N}$\_ICMP}) connections. 

\begin{table}[ht!]
\centering
\scalebox{0.5}{
  \begin{tabular}{|l|l|c|}
  \hline
  \textbf{Category} & \textbf{Feature} & \textbf{Description} \\ \hline \hline
  \multicolumn{3}{|c|}{\textbf{Extracted Features}} \\ \hline
%   & \multicolumn{2}{c|}{Connection Information} \\ \cline{2-3}
  \multirow{4}{*}{\makecell{Connection \\ Information}}   & 1. \textit{$\mathcal{N}$\_conn} & \makecell{Number of connections} \\
  & 2. \textit{$\mathcal{N}$\_normal\_flow} & \makecell{Count of normal connections} \\
  & 3. \textit{$\mathcal{N}$\_back\_flow} & \makecell{Count of background connections} \\ 
  & 4. \textit{$\mathcal{\bar{W}}$\_duration}           & \makecell{Mean of duration of all connection} \\ \cdashline{1-1} \cline{2-3}
%   & \multicolumn{2}{c|}{IP Address} \\ \cline{2-3}
  \multirow{8}{*}{IP Address} & 5. \textit{$\mathcal{N}$\_s\_a\_p\_address}     & \multirow{4}{*}{\makecell{Count of source \\ IP by class}} \\
  & 6. \textit{$\mathcal{N}$\_s\_b\_p\_add}     & \\
  & 7. \textit{$\mathcal{N}$\_s\_c\_p\_add}     & \\
  & 8. \textit{$\mathcal{N}$\_s\_na\_p\_add}    & \\ \cline{2-3}
  & 9. \textit{$\mathcal{N}$\_d\_a\_p\_add}     & \multirow{4}{*}{\makecell{Count of destination \\ IP by class}} \\
  & 10. \textit{$\mathcal{N}$\_d\_b\_p\_add}     & \\
  & 11. \textit{$\mathcal{N}$\_d\_c\_p\_add}     & \\
  & 12. \textit{$\mathcal{N}$\_d\_na\_p\_add}   & \\ \cdashline{1-1} \cline{2-3}
%   & \multicolumn{2}{c|}{Port Information} \\ \cline{2-3}
  \multirow{4}{*}{\makecell{Port \\ Information}}  & 13. \textit{$\mathcal{N}$\_sports>1024}  & \multirow{2}{*}{\makecell{Count of source ports over and below 1024}}  \\
  & 14.  \textit{$\mathcal{N}$\_sports<1024}  & \\
  & 15. \textit{$\mathcal{N}$\_dports>1024} & \multirow{2}{*}{\makecell{Count of destination ports over and below 1024}} \\
  & 16. \textit{$\mathcal{N}$\_dports<1024} & \\ \cdashline{1-1}  \cline{2-3}
%   & \multicolumn{2}{c|}{Protocol Information} \\ \cline{2-3}
  \multirow{4}{*}{\makecell{Protocol \\ Information}} & 17. \textit{$\mathcal{N}$\_icmp}       & \multirow{3}{*}{Count of protocol used} \\
  & 18. \textit{$\mathcal{N}$\_tcp}        & \\
  & 19. \textit{$\mathcal{N}$\_udp}        & \\ \hline
  
  \multicolumn{3}{|c|}{\textbf{Analyzed Features}} \\ \hline
  \multirow{9}{*}{\makecell{Connection \\ Information}}   & 20. \textit{$\mathcal{S}$\_packets}  & \makecell{Entropy - packets transferred} \\
  & 21. \textit{$\mathcal{S}$\_srcbytes }& \makecell{Entropy - bytes from source} \\
  & 22 \textit{$\mathcal{S}$\_bytes }   & \makecell{Entropy - bytes transferred} \\
  & 23. \textit{$\mathcal{S}$\_state} & \makecell{Entropy - connection states}\\
  & 24. \textit{$\mathcal{S}$\_time} & \makecell{Entropy - connection duration} \\
  & 25. \textit{$\sigma$\_time} & \makecell{Standard deviation - duration per connection} \\
  & 26. \textit{$\sigma$\_packets} & \makecell{Standard deviation - packets transferred} \\ 
  & 27 \textit{$\sigma$\_bytes} & \makecell{Standard deviation - bytes transferred} \\
  & 28. \textit{$\sigma$\_srcbytes} & \makecell{Standard deviation - bytes from source transferred} \\ \cdashline{1-1} \cline{2-3}
  \multirow{11}{*}{\makecell{Entropy \\ IP Address}}   & 29. \textit{$\mathcal{S}$\_srcip} & \multirow{2}{*}{\makecell{Entropy - source and \\ destination IP addresses}} \\
  & 30. \textit{$\mathcal{S}$\_dstip} & \\ \cline{2-3}
  & 31. \textit{$\mathcal{S}$\_src\_a\_ip } & \multirow{4}{*}{\makecell{Entropy - source \\ IP by class}} \\
  & 32. \textit{$\mathcal{S}$\_src\_b\_ip}     & \\
  & 33. \textit{$\mathcal{S}$\_src\_c\_ip}     & \\
  & 34. \textit{$\mathcal{S}$\_src\_na\_ip}     & \\ \cline{2-3}
  & 35. \textit{$\mathcal{S}$\_dst\_a\_ip}   & \multirow{4}{*}{\makecell{Entropy - destination \\ IP by class}} \\
  & 36. \textit{$\mathcal{S}$\_dst\_b\_ip}   & \\
  & 37. \textit{$\mathcal{S}$\_dst\_c\_ip}   & \\
  & 38. \textit{$\mathcal{S}$\_dst\_ns\_ip}   & \\ \cline{2-3}
  & 39. \textit{$\mathcal{S}$\_src\_to\_dst} & \makecell{Entropy of source to destination IP \\ with duration and bytes transferred} \\ \cdashline{1-1} \cline{2-3}
  \multirow{6}{*}{\makecell{Entropy \\ Port}} & 40. \textit{$\mathcal{S}$\_srcport} & \multirow{2}{*}{\makecell{Entropy - source \\ and destination ports}} \\
  & 41. \textit{$\mathcal{S}$\_dstport} & \\\cline{2-3}
  & 42. \textit{$\mathcal{S}$\_sports>1024} & \multirow{2}{*}{\makecell{Entropy - source ports}} \\
  & 43. \textit{$\mathcal{S}$\_sports<1024}     & \\ \cline{2-3}
  & 44. \textit{$\mathcal{S}$\_dports>1024} & \multirow{2}{*}{\makecell{Entropy - destination ports}} \\
  & 45. \textit{$\mathcal{S}$\_dports<1024} & \\
  
    \hline
\end{tabular}
}
\caption{Extracted Features - Feature variables from the netflow records aggregated by time window}
\label{tab:features}
\end{table}

The analyzed category develops in-depth feature transformations using the raw values recorded from each of the netflow records (within a time-window). The feature calculations in the category are based on either entropy ($\mathcal{S}$) or standard deviation ($\sigma$) of the features within the time window.

% , where -

% \begin{itemize}
%     \item Entropy: 
%         \begin{equation}
%             \mathcal{S} = - \sum_{i=1}^{n} P(x_{i})log_{2}P(x_{i})
%         \end{equation}
%     \item Standard Deviation:
%         \begin{equation}
%             \sigma = \sqrt\frac{\sum_{i=1}^{n}(x_i - \bar{x})^2}{n - 1}
%         \end{equation}
% \end{itemize}

% where, $n$ is the number of connections in the time-window, $x_{i}$ is the observed feature value, $P(x_{i})$ is the probability mass function of $x_{i}$, and $\bar{x}$ is the mean of the feature. 

Similar to the extracted features, the analyzed features were also developed for connection information, IP address, and port data. Connection based entropy of packets transferred (\textit{$\mathcal{S}$\_packets}), the total number of bytes (\textit{$\mathcal{S}$\_bytes}), bytes from source (\textit{$\mathcal{S}$\_srcbytes}), connection states (\textit{$\mathcal{S}$\_state}), and duration (\textit{$\mathcal{S}$\_time}) were recorded for each time window. For each of the entropy values, standard deviations were also measured (excluding the state entropy) - \textit{$\sigma$\_packets}, \textit{$\sigma$\_bytes}, \textit{$\sigma$\_srcbytes}, and \textit{$\sigma$\_time}. Entropy in IP address were recorded for change within different classes for both source and destination. Apart from the standard entropy calculation for the IP addresses, \textit{$\mathcal{S}$\_src\_to\_dest} was developed to map single source to destination connections along with their connection duration and amount of data transferred in each connection. Changes in port utilization within a time-window was also captured as entropy variables for source and destination and utilization standard versus unknown ports.

The final output of the feature extraction stage represents the following data:

\begin{equation}
    f_{v}^{i} = {<x_{1}^{i} \cdots x_{47}^{i}>} : y^{i} = l^{i}
\end{equation}

Where, $f_{v}^{i}$ is the feature vector for the window $i$, $l^{i}$ is the label of the window represented by $y^{i}$, where
\begin{equation}
l^{i} = \left\{
  \begin{array}{lr}
    y^{i} : 0 \text{ if attack flows in window($i$) $= 0$}\\
    y^{i} : 1 \text{ if attack flows in window($i$) $\ge 1$}\\
  \end{array}
\right.
\end{equation}

\subsection{Building Machine Learning Models}
\label{sec:mlmodel}

The machine learning algorithms explored in here are - 1) Random Forest and 2) Multi-Layer Perceptron (Deep Learning). A Random forest model can be defined as $h = \{h_{1}(X), \cdots, h_{j}(X)\}$, where an ensemble of decision tree classifiers ($h_{j}(X)$, with $j$ trees) are used to make collective decisions. $X$ is the model training data of the features vectors ($f_{v}^{i}$) and their corresponding labels ($l^i$).

Multi-Layer Perceptrons (MLP) \cite{rumelhart1985learning, rosenblatt1958perceptron} use feed-forward artificial neural networks consisting of an input layer, multiple hidden layers, and an output layer. Each layer is made up of perceptrons which compute a single output ($\gamma$) based on a input vector  defined by $\gamma = \varphi(\sum_{i=1}^{n} w_{i} x_i + b)$, where $w$ denotes the vector of weights for the perceptron, $x$ is input vector (either from the actual input vector or previous layer outputs), $b$ being the bias, and $\varphi$ the activation function. The weights and biases are adjusted in the perceptrons in each layer based on the training input vector ($f_{v}^{i}$) and their corresponding labels ($l_i$) using back and forward propagation.

For each of the models, the entire dataset was separated into training data (70\% of the dataset) for model development, and test data (remaining 30\%) for validation. For the Random forest, the number of estimators/decision trees used in the initial model was $j = 10$, maximum of 6 features per estimator, no maximum depth for the trees, and requires two samples to split an internal node. The MLP model, was based on 4-layer approach, with one input, 2 hidden layers each followed by a dropout layer of 50\% dropout (to prevent overfitting), and one output layer. Each layer utilizes Rectified Linear Unit (ReLU) as its activation function $\varphi$, and binary cross-entropy for the loss function. Performance of each model was recorded for their accuracy (overall classification), precision ($\frac{tp}{tp+fp}$), recall ($\frac{tp}{tp+fn}$), and F-1 score ($\frac{2.precision*recall}{precision+recall}$), where $tp$ is the number of true positive classified of attack labels, $fp$ is the number of false positives classified on attack labels, $fn$ is the number of false negatives classified by the models.

% we surveyed Random Forest and Deep Learning models for detecting botnets. We trained each algorithm on aggregated binetflow files based on the attack that was present in each file. Each file was aggregated by one second intervals and then all the intervals from each file were concatenated into one featureset. We excluded background connections from the dataset and only looked at Normal and Botnet connections. Background flows would often drown out other connections and so excluding them allowed a more balanced dataset. The featureset was split into 70\% training data and 30\% testing data. We then compared the performance of each algorithm with respect to their accuracy, precision, and recall. In addition to detecting attack type, we are also including a dataset that combines all binetflows and tries to classify each connection into the Botnet type that was used. There was seven types of bots: Neris, Rbot, Virut, Sogou, Murlo, NSIS.ay, and Menti. 

% We surveyed the following machine learning algorithms; SVM, Random Forest, Decision Trees, Naive Bayes, and Deep Learning. We trained each algorithm on the individual binet flow files from the CTU-13-Dataset to produce models for predicting attack traffic. Each file was slit into 70\% training data and 30\% testing data. We then compared the performance of each algorithm with respect to their accuracy, precision, and recall. We were able to see the impact that the unbalanced files had on the models, as the precision and recall on those files was low.

% \noindent
% \textit{Window Size Tuning}: 

Different window sizes for the feature vectors ($f_{v}^{i}$) in the data were explored while evaluating the performance of the model. Window sizes ranging from 0.1 all the way with 60 second intervals were evaluated for different types of attacks/communication (DDoS, SPAM, and IRC) with the deep learning and random forest algorithms. For each of these intervals the F1-Score was recorded for cross-evaluation. An exhaustive search was also conducted for evaluation of hyper-parameter values and architecture layouts which yield better performance of detecting attack windows in the learned models. In the random forest models, a grid search method was used, where multiple estimators (10 - 700) were evaluated for their performance. Other parameter values such as min-split of leaves, max depth of the estimators, and number of features per estimators were also evaluated for their F-1 score. In the MLP model, multiple architectures with different number of hidden layers, number of perceptrons per layer, and different activation functions were tested (Tanh, Sigmoid, LeakyReLU) were tested for their accuracy.

Each of the models were also run through a k-fold cross validation ($k=10$) with random selection of training and test datasets, where the mean scores of the models were taken into account for performance metric comparisons.

% Each of the high performing algorithms could be modified such that it could yield higher accuracy, precision, and recall (e.g.\ changing number of estimators in Random Forest). We used a grid search algorithm to tune the configurable parameters of these algorithms
% and give the best results. For Random Forest, it seemed that 100 trees was optimal amount and using entropy as our criterion gave the best results. We also had Random Forest use all features in it's trees.
% Decision Trees only needed an entropy criterion. Deep learning used a Multi-Layer Perception (MLP) model which consisted of five layers, two dropout layers with 0.5 value, and two dense layers with 64 nodes each, the output layer used binary cross entropy.

\section{Results and Analysis}
\label{results}

% \begin{table}
% \scalebox{0.5}{
%   \noindent\resizebox{\textwidth}{!}{
%   \begin{tabular}{| c | c | c |}
% \hline

% \multirow{2}{*}{\textbf{\makecell{Attack \\ Type}}} & \textbf{Random Forest} & \textbf{Deep learning (MLP)} \\ \cline{2-3}
% & \textit{Accuracy - Precision - Recall - F1 Score} & \textit{Accuracy - Precision - Recall - F1 Score} \\ \hline
% DDOS        & 0.9764 - 0.9793 - 0.9329 - 0.9708 & 0.7835 - 0.9837 - 0.2652 - 0.3431 \\ \hline
% SPAM        & 0.9905 - 0.9913 - 0.9927 - 0.9942 & 0.9298 - 0.9219 - 0.9586 - 0.9399 \\ \hline
% IRC         & 0.9862 - 0.9864 - 0.9856 - 0.9912 & 0.9169 - 0.9320 - 0.9027 - 0.9171 \\ \hline
% Combined    & 0.9975 - 0.9977 - 0.9981 - 0.9923 & 0.9580 - 0.9359 - 0.9961 - 0.9435 \\ \hline

%   \end{tabular}
%     }
%     }
%   \caption{Initial Model Performance Metrics - Performance scores (Precision, Recall, F1-score) of the three different types of attacks and combined features based on one second window size and default model parameters.}
% \label{tab:model_scores}
% \end{table}

\subsection{Initial Results}
Table \ref{tab:final_metrics} describes the overall performance of the Random Forest and Deep Learning (MLP) models. The reported values shown in the table are for base/initial models with single window size for all attacks (1 sec) and default model parameters as described in Section \ref{sec:mlmodel}. In the case of Multi-Layer Perceptron model, where a binary classifier is used to predict windows with attack/non-attack features, the resulting model is able to perform with an overall F1-score (value between 0 - 1) of \textit{.34} in DDoS, \textit{.93} in SPAM, and \textit{.91} in IRC data. The model performs very poorly on the DDoS data indicated by low recall score of \textit{.26}, suggesting its inability to correctly identify attack windows. In case of SPAM and IRC, the model is able to perform well with good accuracy score. 

% Merging all the data (DDoS, SPAM, and IRC) into one dataset, improves the model's F1-Score to \textit{.94}. The increase in accuracy suggests lack of large amount of training data in the initial datasets (DDoS). Another possible issue could be an inaccurate window size (default is 1 sec) for our features, which does not capture a representative feature set of a DDOS connection.

In comparison, Random Forest consistently performs well on each of the data with F1-scores of \textit{.97} in DDoS, \textit{.99} in SPAM, and \textit{.98} in IRC data. While the accuracy of the model is high, we can notice the performance of the model for DDoS attack is lower when compared others. These issues lead to a further investigation of the models on their window sizes for each of the datasets.

\begin{figure}[ht!]
  \centering
  \scalebox{0.75}{
  \includegraphics[height=2.5in]{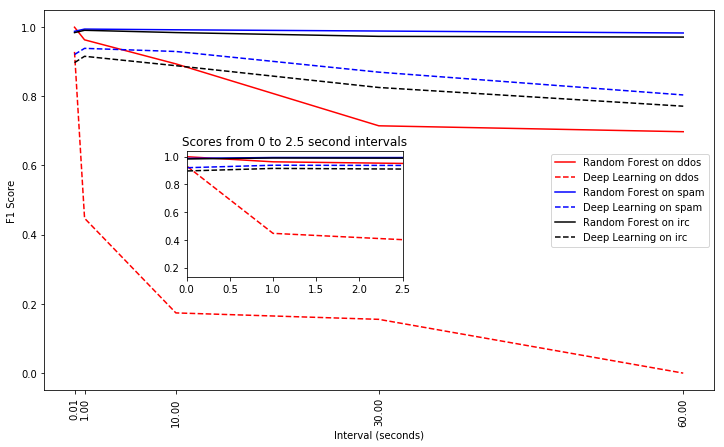}}
  \caption{Window Size versus F1-score: Analysis of model performance in different window size aggregation.}
  \label{fig:interval}
  \vspace{-1em}
\end{figure}

% TODO: Histograms, analysis of raw data.
% TODO: Describe machine learning models.
% TODO: Explain precision/recall ROC. (Before feature tuning)

\subsection{Model/Feature Tuning}
Figure \ref{fig:interval} describes the performance (mean F1-score) of the models in different scenarios of attacks (DDoS, SPAM, and IRC) with different window sizes for aggregation of features. In case of DDoS, the both Random Forest and MLP have higher F1-scores with lower window size (in initial model was 1 sec) of 0.01 second. Increasing the time windows to be greater than 0.01 seconds, leads to significant degradation in DDoS detection performance. This can be attributed to lower time duration of individual DDoS connections, where smaller window size will lead to accurate representation of attack features, and in turn lead to better model performance. For the SPAM and IRC datasets, the analysis did not show any benefits of changing the default 1 second interval for window sizes, instead lead to inferior performance (drops in F1-score) at lower (0.01) and higher (10, 30, and 60 second) time windows.

Analysis of the individual model hyper-parameters did not show any significant increase in performance in both the algorithms. In Random Forest increase in the number of estimators showed only 0.01 gain in model accuracy. Similar non-significant increase in model performance was observed in MLP, where larger networks of perceptrons and layers were used. The models were analyzed for the reported prediction probability values instead of just the default to  detect all attack windows. The decision threshold of detecting attack window was decreased to 0.3 (based on the Recall scores), where window with predicted probability of attack is $1 \ge 0.3$ is taken to be an attack window, and in turn classifies a window to be normal if predicted probability of $0 \le 0.7$. While this decreases the F-1 score of the normal windows ($0$), it aims for increased accuracy in detecting attack windows $1$.

\subsection{Final metrics}

% \begin{itemize}
%     \item Confusion Matrix %model tuning
%     \item ROC %model tuning
%     \item Feature Importance %model tuning
% \end{itemize}

\begin{table}[ht!]
\scalebox{0.5}{
  \noindent\resizebox{\textwidth}{!}{
    \begin{tabular}{| c | c | c |} \hline

    \multirow{3}{*}{\textbf{\makecell{Attack \\ Type}}} & \textbf{Random Forest} & \textbf{Deep learning} \\ \cline{2-3}
    & \textit{Accuracy - Precision - Recall - F1} & \textit{Accuracy - Precision - Recall - F1} \\ \cline{2-3}
    & \multicolumn{2}{|c|}{\textbf{Initial Untuned Model}} \\ \hline
    DDOS        & 0.9764 - 0.9793 - 0.9329 - 0.9708 & 0.7835 - 0.9837 - 0.2652 - 0.3431 \\ \hline
    SPAM        & 0.9905 - 0.9913 - 0.9927 - 0.9942 & 0.9298 - 0.9219 - 0.9586 - 0.9399 \\ \hline
    IRC         & 0.9862 - 0.9864 - 0.9856 - 0.9912 & 0.9169 - 0.9320 - 0.9027 - 0.9171 \\ \hline
    % Combined    & 0.9975 - 0.9977 - 0.9981 - 0.9923 & 0.9580 - 0.9359 - 0.9961 - 0.9435 \\ \hline
    & \multicolumn{2}{|c|}{\textbf{Tuned Model}} \\ \hline
    DDOS        & 0.999 - 0.999 - 0.999 - 0.999 & 0.942 - 0.946 - 0.965 - 0.956 \\ \hline
    SPAM        & 0.993 - 0.993 - 0.996 - 0.994 & 0.907 - 0.962 - 0.879 - 0.918\\ \hline
    IRC         & 0.992 - 0.989 - 0.995 - 0.992 & 0.895 - 0.925 - 0.857 - 0.889\\ \hline
    % Combined         & 0.990 - 0.989 - 0.994 - 0.991 & 0.918 - 0.927 - 0.933 - 0.930\\ \hline
    
      \end{tabular}
      }
      }
  \caption{Performance Metrics of Models: Accuracy, Precision, Recall, and F1-score of initial/untuned versus tuned models}
    \label{tab:final_metrics}
\end{table}

\begin{figure}[ht!]
  \centering
  \includegraphics[width=\linewidth]{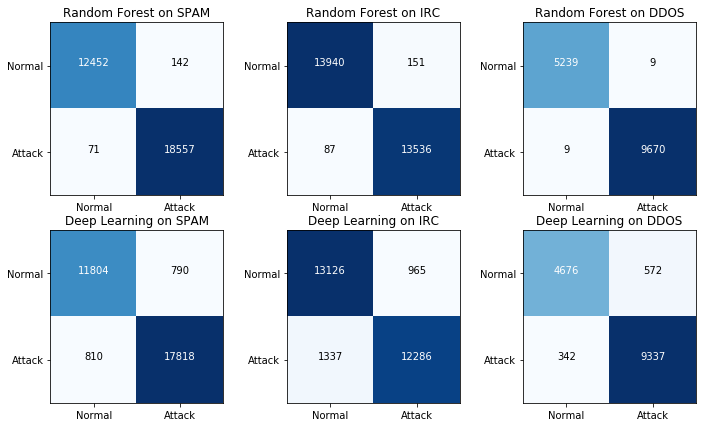}
  \caption{Confusion Matrices Of Each Model On Each Attack: Prediction matrix of Deep Learning and Random Forest with tuned parameters.}
  \label{fig:rf_confusion}
  \vspace{-1em}
\end{figure}

Table \ref{tab:final_metrics} outlines the performance metrics of tuned models. With the window size tuning we see improvements in performance across all models (as compared to initial values), especially in the DDoS data. The MLP models gained a significant boost in their performance in detecting DDoS attack from an F1-score of \textit{0.34} in initial models to \textit{0.95} in 10 millisecond window size aggregation. As a result, the combined score of the MLP model also improves to $0.93$.

In the case of Random Forest, the models perform on nearly perfect accuracy across all types of attacks. In the case of DDoS we observe an increase in accuracy by 2\% where F1-Score increase from \textit{0.97} to \textit{0.99}. With the probability tuning, Figure \ref{fig:rf_confusion} shows the accuracy of Random forest in detecting attack windows with only 8 windows mis-classified as normal while getting 9,671 windows accurately detected (overall 17 windows incorrectly classified out of14,927 ) in training data. Similar results were also observed in detecting SPAM and IRC windows.

% Earlier in this paper we described our methodology of tuning the parameters of the machine learning algorithms in order to get higher performing results. After tuning the parameters, we see some improvements on all metrics from each algorithm. Looking at the final metrics in Table \ref{final_metrics} we immediately notice the effects on changing the time interval in which we aggregated connections for DDOS. When initially we used one second, we found better results aggregating on ten millisecond intervals instead. Random Forest does nearly perfect on DDOS now and Deep Learning jumped from 70\% to almost 95\%. The confusion matrix for Random Forest is shown in Figure \ref{rf_confusion} since Random Forest seems to do especially well in detecting botnets in DDOS attacks.

\section{Discussion}
\subsection{Feature Tuning}
Analysis of performance metrics observed for feature (window) and model tuning suggests large gains in model detection accuracy can be observed by using different window sizes for feature aggregation. In comparison, model hyper-parameter tuning provide negligible performance benefits, and can in some cases lead to increase in complexity of the underlying model. Our analysis suggests using 0.01 sec window size for DDoS, and 1 sec windows for SPAM and IRC, in order to get the best performance out of the trained models. In a real-world scenario, this would involve training multiple models with different aggregated window size based features. A general model which uses a single aggregation size will not be suitable for detection of every type of attack, as shown by our initial results.

% Comparison of performance in initial/untuned versus final/tuned models shows the importance of evaluating the window size, model parameters, and probability tuning for each attack type. One of the biggest gains in the performance scores was from the change in aggregation window size in DDoS. In a real-world scenario, this would involve training multiple models with different aggregated window size based features. A general model which uses a single aggregation size will not be able to perform well for every type of attack. This was highlighted in the window size analysis, where DDoS was performing poorly in $\ge 0.01$ sec aggregation, while SPAM and IRC were performing better at 1.0 sec. A compromise between complexity of using multiple models (with different window based features) and the simplicity of a single window based approach, can be by using $0.01$ window size for model development, where we do not see a large drop in accuracy of SPAM / IRC, but are able to get the increased performance in DDoS detection.

\begin{figure}[ht!]
  \centering
%     \scalebox{0.75}{
  \includegraphics[width=\linewidth]{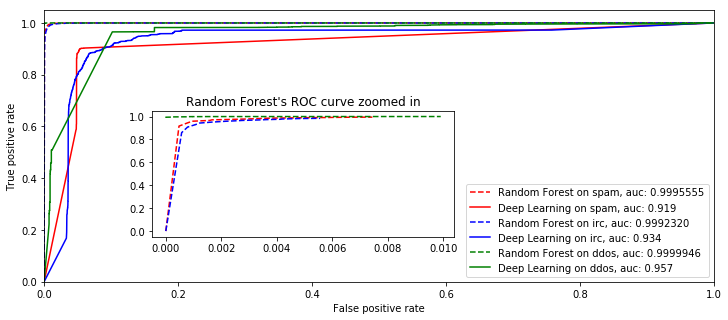}
%   }
  \caption{Multi-Layer Perceptron \& Random Forest ROC curve: ROC curve for each attack type (tuned model).}
  \label{fig:mlp_roc_curves}
  \vspace{-1em}
\end{figure}

% \begin{figure}
%   \centering
%   \scalebox{0.75}{
%   \includegraphics[height=2.5in]{./images/roc_updated.png}
%   }
%   \caption{Random Forest ROC curve: ROC curve for each attack type (tuned model).}
%   \label{fig:rfroc_curves}
% \end{figure}

\subsection{Model Performance}
Based on the performance metrics of the tuned models, it can clearly be observed that Random Forest based models outperform MLP models in detecting attack windows in every category. This is especially highlighted in the model Receiver Operating Characteristics (ROC) curves outlined in Figure \ref{fig:mlp_roc_curves}. While, the Area Under the Curve (AUC) values for MLP are above \textit{0.90} for all types of attacks, the model performs the worst on detection of SPAM attacks. MLP models work well in detecting DDoS attack windows with AUC score of \textit{0.97}, closely followed by botnet communication detection of IRC at \textit{0.95}. In comparison, the Random Forest models perform at near perfect accuracy with AUC scores of DDoS - \textit{1.0}, IRC - \textit{0.99}, and SPAM - \textit{1.0}. 

The relative poor performance of MLP models can partially be attributed to the number of available training samples. MLP models (and more generally Deep Learning) greatly benefit from large amounts of training data and can outperform traditional models in such a case. This was also observed in detecting DDoS attacks, where 10 millisecond window size aggregation lead to increased model performance. While dataset used in our experiments were limited, in large organizations this would note be the case, and MLP based methods can be used for detection of attacks. In scenarios, where amount of labeled data is limited ensemble based methods such a Random Forests will outperform Deep-Learning based approaches as shown by our results. 

\begin{figure}[ht!]
  \centering
  \scalebox{0.85}{
  \includegraphics[height=1.1in]{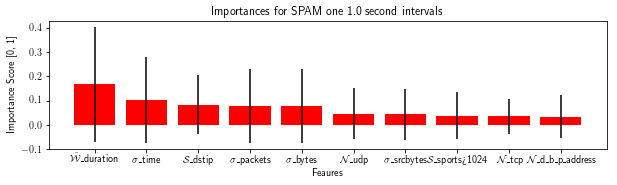}
  }
  \scalebox{0.85}{
  \includegraphics[height=1.1in]{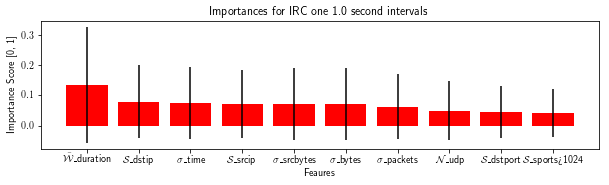}
  }
  \scalebox{0.85}{
  \includegraphics[height=1.1in]{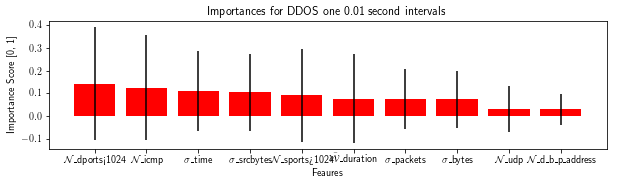}
  }
  \caption{Feature Importance Random Forest - Importance of features in different types of attack with normalized scores (0 - 1)}
  \label{fig:rf_feature_importance}
  \vspace{-1em}
\end{figure}

\subsection{Feature Importance}
Diving deeper into the characteristics of the Random Forest model, Figure \ref{fig:rf_feature_importance} shows the importance of top ten features in each attack category. In the case of SPAM, the mean duration of the connections ($\mathcal{\bar{W}}$\_duration) play an important role in detecting attacks, followed by standard deviation of connection duration ($\sigma$\_time), entropy of destination IP address ($\mathcal{S}$\_time) suggesting SPAM messages sent out to large number of destinations, standard deviation of packets ($\sigma$\_packets) - difference in the number of packets in attack versus normal windows, and bytes ($\sigma$\_bytes) transferred within a time window where SPAM communication will see a larger amount transferred. While other features (feature number - 19, 21, 42, 18, 10) play a role in detection of the SPAM based attacks they are of less significance from the top five features. 

Similarly, in IRC detection, the mean duration of the connections ($\mathcal{\bar{W}}$\_duration) again is of significant importance followed by entropy of destination IP address ($\mathcal{S}$\_destip) suggesting a controller communicating with a large number of bots, standard deviation of connection duration ($\sigma$\_time) where communication windows have smaller connection durations, entropy of source IP address ($\mathcal{S}$\_srcip) - a few/single controller communicating to bots, standard deviation of bytes ($\sigma$\_bytes), and packets ($\sigma$\_packets) transferred in communication during attacks. $\mathcal{N}$\_udp also plays a role as UDP is one of the preferred modes of communication in IRC. 

In DDoS attack scenario, the models rely on identifying the windows where communication exists destination ports $\ge 1024$ ($\mathcal{N}$\_dports>1024). Also, in DDoS based attacks the number of ICMP based flows increases as identified by the $\mathcal{N}$\_icmp feature importance. Standard deviation of flow duration ($\sigma$\_time) - short flows for DDoS, amount of source bytes transferred ($\mathcal{N}$\_srcbytes), number of source ports greater than 1024 ($\mathcal{N}$\_sports>1024), mean duration of flows ($\mathcal{\bar{W}}$\_duration) , and standard deviation of packets ($\sigma$\_packets) - bytes ($\sigma$\_bytes) play key roles in identifying attack windows.

The analysis suggests the models can perform accurate detection of the attack/communication windows with limited number of features (in comparison to the previously identified 45 features), which be used to store lower number of variables thereby considerably decreasing the size of the feature set, and in turn reducing the complexity of the machine learned models.

\section{Conclusion}
\label{sec:conclusion}

% What did we learn 
% What are our limitations - annotated data Subject Matter Expert who can annotate the data. need for large data in MLP, low data - Random forest, we did not take into account background data - only works when we have labels for both attack and non attack flows
% benefits: low data - Random forest, can use 0.01 msec with adequate performance, we can built with low number of features not 45, models are really accurate. 
% Future work - test lower number of features versus accuracy, general model applicable to all attacks,  we how this model does with background, is the model transferable,

We analyzed the capabilities and limitations of two key machine learning models in their ability to detect attacks in performed over network communication. Using Big Data approaches to develop features (from large datasets) and train - test machine learning models, we were able to identify operating characteristics of the learned models in real-world scenario. Our analysis of the models found feature engineering to be a key contributor to improving machine learning models, where the biggest increase in model accuracy resulted from changes in window sizes. Exploration of hyper-parameters related to the model did not show any significant increase both Random Forest and MLP models. Comparing the models itself, we observe Random Forest is capable of operating at near perfect accuracy in detecting different types of botnet attacks/communication when compared to MLP models. The limitation of annotated data plays a key role in the lower performance of MLP models, so in scenarios where there is availability of large annotated data (using honey-pots, subject matter expert, detection over time) MLP based models could potentially be effective. However, in scenarios where there is limited availability of such annotated data, Random Forest models will outperform their counterparts. 

The study also observed multiple aggregation windows sizes perform the best for different types of attacks. While this is negates the utilization of a generalized model, in case of resource constraints (space and performance issues related to multiple datasets and models), a generalized dataset using 0.01 second aggregation can be used with best performance in DDoS attacks, and negligible degradation in accuracy for SPAM and IRC. In order to further solve the dataset / model complexity, lower number of features can also be used for model training as identified by out feature importance analysis. 

Future efforts include analyzing the performance of models when we have unlabeled background traffic and/or highly imbalanced data. We also plan to evaluate of other types of attacks/communication such as peer-to-peer, click-fraud, and other botnet/malware related network data, and use the metrics for comparison to standard detection tools (such as Bot-Sniffer \cite{gu2008botsniffer}).  

% From our analysis, we found success in detecting botnet attacks using our method of aggregating connections. We notice that Random Forest was especially effective in classifying this data, reaching to almost 99\% accuracy, precision and recall. Deep Learning was effective as well but just couldn't match Random Forest's performance. However, this accuracy was achieved by using only normal and attack connections, excluding background connections. We can expect that when we add background, we may lose some accuracy. 

% We found an increase in DDOS detection when we lowered the time interval because it increased the change of having intervals with only normal or only attack connections. Adding background connections may drown out some of the features of attack and normal connections, thus confusing the machine learning models. We see limitations in that our models may have trouble in finding attack connections among different types of connections. 

% In the future, we can explore how we can improve the detection of botnet attacks when background connections are included. We know when we isolate normal and attack connections, the differences are clear enough for machine learning models to clearly classify. Having machine learning models that are aware of these differences could help -

\section{Acknowledgement}
\vspace{-0.5em}
This research was supported by University of North Carolina - Greensboro new faculty start-up funds. 

% \nocite{*} %make all source appear even if they aren't cited in the body.

\bibliography{bib}
\bibliographystyle{IEEEtran}

\end{document}